\documentclass[preprint,showpacs,floats,letterpaper,floatfix,,groupedaddress,eqsecnum]{revtex4}

\usepackage{subfigure}

\usepackage{amssymb,amsmath}
\usepackage[dvips]{graphicx}
\usepackage{csquotes}

\usepackage{dcolumn,epsfig}
\usepackage[left]{lineno}

\usepackage[final]{pdfpages}

\begin{document}

\title{An analog simulation experiment to study free oscillations of a damped simple pendulum}

\author{Ivan Skhem Sawkmie$^1$, and Mangal C. Mahato$^1$$^,$}
\email{mangal@nehu.ac.in}
\affiliation{$^1$Department of Physics, North-Eastern Hill University,
Shillong-793022, India}

\begin{abstract}
The characteristics of drive-free oscillations of a damped simple pendulum under sinusoidal potential force field differ from those of the damped harmonic oscillations. The frequency of oscillation of a large amplitude simple pendulum decreases with increasing amplitude. Many prototype mechanical simple pendulum have been fabricated with precision and studied earlier in view of introducing them in undergraduate physics laboratories. However, fabrication and maintenance of such mechanical pendulum require special skill. In this work, we set up an analog electronic simulation experiment to serve the purpose of studying the force-free oscillations of a damped simple pendulum. We present the details of the setup and some typical results of our experiment. The experiment is simple enough to implement in undergraduate physics laboratories.

\end{abstract}
\vspace{0.5cm}
\date{\today}

\maketitle

\section{Introduction}
The determination of the period of a pendulum is a common physics experiment performed in schools and junior colleges. Using the small amplitude approximation, the amplitude-independent period of oscillation of the pendulum is related to the acceleration due to gravity in the laboratory. Although the air resistance ultimately brings the pendulum to a stop, its effect on the period of oscillation is small and hence usually ignored. An equivalent experiment is also performed using an LCR circuit. The voltage oscillation is initialized with an {\it{ac}} voltage drive and then the drive is switched off at an appropriate time to observe decaying voltage oscillations in time using an oscilloscope. However, these are essentially damped harmonic oscillations.

An experiment in which the amplitude of oscillation of the pendulum is large is an advanced step. The sinusoidal force experienced by this simple pendulum (derived from the gravitational potential) cannot be approximated to be harmonic. The period of oscillation of the simple pendulum depends on its amplitude. In the undamped case the period is given in terms of elliptic integrals\cite{Sommerfeld}. The effect of air resistance and other kinds of dampings have also been investigated theoretically\cite{Johannessen} and experimentally in real mechanical simple pendulum\cite{Blackburn1989,Blackburn1998,Patrick}. The damping force could depend linearly (Stokesian), and/or quadratically on velocity. The mechanical fulcrum could also contribute to velocity independent damping\cite{Patrick}. There is a large number of similar investigations on the subject as reported in this journal (Am. J. Phys.) as well as elsewhere \cite {Fulcher, Ganley, Cadwell, Molina, Kidd, Millet, Parwani, Hite, Belendez2009, Turkyilmazoglu2010, Turkyilmazoglu2011, Belendez2011}. However, introducing mechanical simple pendulum in college and university teaching laboratories may not be as simple for it requires special mechanical skills and infrastructure. In this work we propose an electronic circuit equivalent of the underdamped simple pendulum experiment. The experimental setup is simple enough to fabricate, implement and maintain with relatively small expenditure in any undergraduate teaching laboratory.

The study of simple pendulum has pedagogic relevance because it has exact analogies in, and can be considered as the prototype of, many other phenomena of physical interest \cite{Risken, Kleppner}. Apart from a particle 
moving on the surface of a sinusoidal potential \cite{Saikia,arxiv}, some of the examples being the experimental study of the chaotic motion of a mechanical pendulum \cite{Blackburn1989,Blackburn1998},
ionic motion in a superionic conductor, and a phase motion in a Josephson junction \cite{Falco,Barone}. 
The structure of the equations of motion for these phenomena is identical only 
the physical significance of the parameters differ.

The equation of motion of a simple pendulum is exactly equivalent to a particle moving in a medium of friction 
coefficient $\gamma$ along a potential $V(x)=-V_0\cos(kx)$ and driven by the external periodic force $F(t)=
F_0\sin(\omega t)$. The equation of motion of the pendulum (identifying $x$ with the angular displacement $\theta$, etc) is thus given by
\begin{equation}
m\frac{d^{2}x}{dt^{2}}=-\gamma\frac{dx}{dt}-V_0k \sin(kx)+F_0\sin(\omega t).
\label{eqn1}
\end{equation}

Here the system is considered to be underdamped $(\gamma<<2\omega_0)$ where $\omega_0=\sqrt{\frac{kV_0}{m}}$ is the natural frequency of free oscillation (at small amplitude). The driven pendulum, when the damping is small, has so far not found exact 
analytical description \cite{Kittel}. Therefore, it is quite educative to experimentally study the free ($F_0=0$) oscillation of a simple pendulum even if only in an equivalent electronic circuit.

The dimensionless form of the equation of motion is:
\begin{equation}
\frac{d^{2}x}{dt^{2}}=-\gamma\frac{dx}{dt}-\sin(x)+F_0\sin(\omega t).
\label{eqn2}
\end{equation}

Here, all parameters are dimensionless and written by taking $m,~k$, and $V_0$ as independent parameters and setting $m=k=V_0=1$. Note that, in these dimensionless units, the natural (angular) frequency $\omega_0$ turns out to be 1. The free oscillation of a damped pendulum is described by setting $F_0=0$. We expect the amplitude of the free but damped oscillation to decay as $Ae^{-\frac{\gamma }{2}t}$ and also the frequency of oscillation to decrease with amplitude\cite{Johannessen} unlike in an LCR circuit where the frequency $\omega _1=\sqrt{\omega _0^2-\frac{\gamma ^2}{4}}$ is independent of amplitude.

In the following, we give the details of the experimental setup. A brief explanation of a similar experimental setup can also be found in Ref. \cite{arxiv}. We then present the results from our experiment in graphical form. We also provide a brief discussion on the results.

\maketitle
\section{Experimental Setup}

Our experimental setup is shown in Fig. 1 in a block diagram form. The sinusoidal input voltage used in our experiment is taken from the Agilent 33500B series waveform generator. The waveforms are recorded using the  InfiniiVision MSO-X 3014A oscilloscope from Agilent Technologies. However, any reasonable waveform generator and oscilloscope (for example, Keysight InfiniiVision 1000 X-Series DSO (50MHz, 2 Ch)) can be used to carry out the experiment.

The electronic circuit shown in Fig. \ref{fig1} is designed and set up to simulate an equation, given below, similar to the equation of motion given by Eqn. (\ref{eqn1}) or its dimensionless form Eqn. (\ref{eqn2}). Here, the system is initially driven periodically by an external periodic input current $I_{inp}(t)=I_0\sin(\omega t)$, derived from an input voltage $V_{in}(t)$. However, to obtain the free oscillation we finally switch off the drive. 

The Kirchhoff condition at A, as in Eqn. (\ref{eqn1}), is given by the equation:
\begin{equation}
R_2C_1C_2\frac{d^{2}V_{out}}{dt^{2}}=-\left\{ \frac{R_2C_1}{R_B}+C_3+\frac{R_2C_2}{R_A} \right\} \frac{dV_{out}}{dt}-\frac{U_0}{V_0}
\sin\left(\frac{V_{out}}{V_0}\right)+\frac{V_{in}(t)}{R_1}
\label{eqn3}
\end{equation}

\begin{figure}[htp]
\centering
\includegraphics[width=14cm,height=6cm]{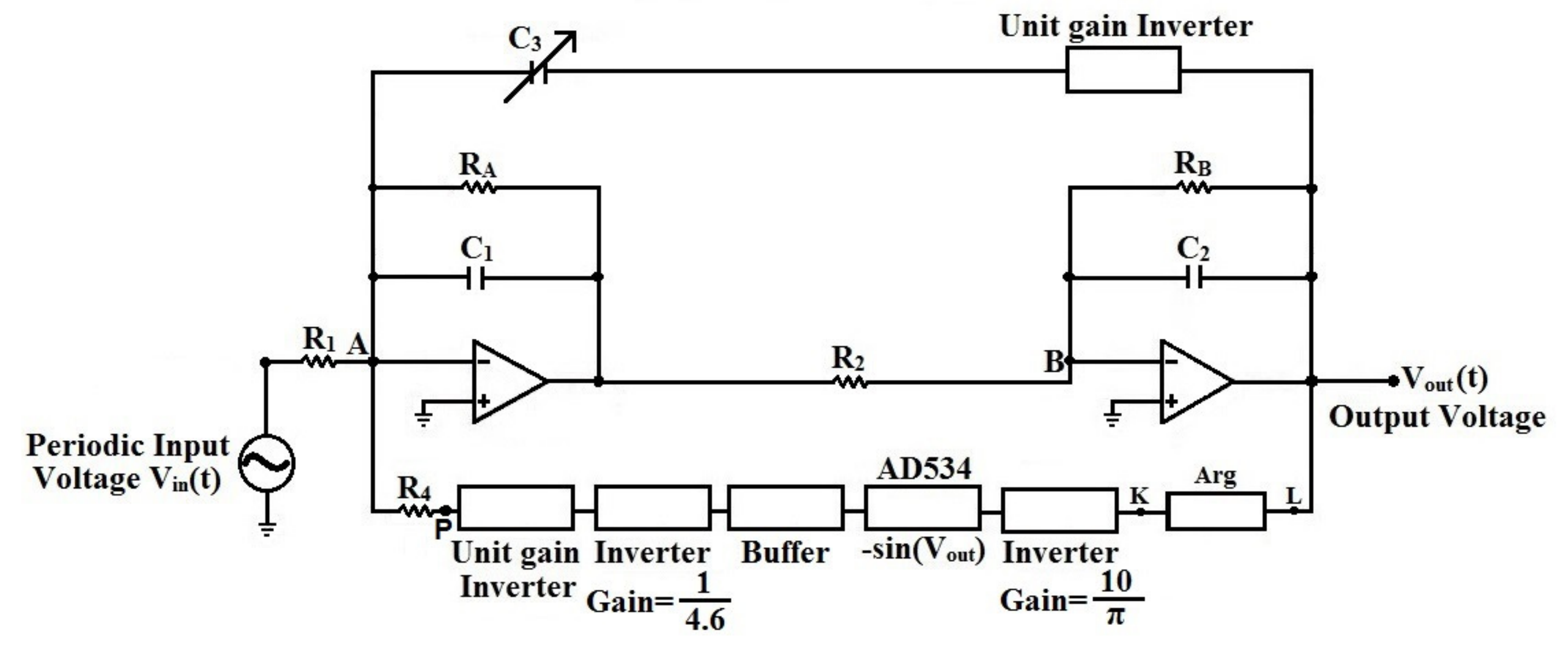}
\caption{Block diagram to simulate Eqn. (\ref{eqn3}). Here the parameters $R_1=5.1K\Omega$ when $V_{in}^0\approx$ 202mVpp, $R_2=5.1K\Omega$, $R_4=10.0K\Omega$, $R_A=1M\Omega$, $R_B=470K\Omega$, $C_1=1.0nF$ and $C_2=10.0nF$ are fixed parameters whereas $C_3$ is a variable parameter (for example, $C_3$ is set equal to $212.47pF$ for $\gamma =0.1181$). The block Arg is elaborated in Fig. \ref{fig2}.}
\label{fig1}
\end{figure}

Here we have taken $V_0=1$volt, $U_0=\frac{V_{0}^2}{R_4}$volt$^2$/ohm, $m=R_2C_1C_2$ A volt$^{-1}$sec$^2$, $k=\frac{1}{V_0}$ volt$^{-1}$ and $V_{in}(t)=V_{in}^0\sin(\omega t)$ volt, where $V_{in}^0$ is the amplitude of the input (signal) voltage, $\omega=2\pi f$ ($f$ is the frequency of the periodic input current). Eq. (\ref{eqn3}) is written in dimensionless units \cite{desloge} by setting the parameters $m=1$, $U_0=1$ and $k=1$. The equation with reduced variables denoted again by the same symbols, corresponding to Eq. (\ref{eqn3}) is written as
\begin{equation}
\frac{d^{2}V_{out}}{dt^{2}}=-\left\{\frac{R_2C_1}{R_B}+C_3+\frac{R_2C_2}{R_A}\right\}\frac{dV_{out}}{dt}-
\sin(V_{out})+I_0\sin(\omega t)
\label{eqn4}
\end{equation}

Comparing Eqs. (\ref{eqn4}) and (\ref{eqn2}), we see that these two equations are similar with damping coefficient $\gamma =\{\frac{R_2C_1}{R_B}+C_3+\frac {R_2C_2}{R_A}\}(R_4/m)^{0.5}$ and $I_0=\frac {V_{in}^0R_4}{R_1V_0}$. Changing the value of $C_3$ changes the value of $\gamma$ since other terms contributing to the value of $\gamma$ are kept fixed. In our experiment, the output voltage $V_{out}(t)$ is analogous to the trajectory $x(t)$ of Eqn. (\ref{eqn2}).

For convenience, we give in, Table \ref{table1}, the relationship between the dimensionless units and dimensioned units for a few physical quantities.

\begin{center}
\begin{table*}[htp]
\caption[Table:] {This table shows the relationship between the value in dimensioned and dimensionless units. The parameters $m$, $U_0$ and $k$ are as defined in the text with the parameter values given in the caption of Fig. \ref{fig1}.}
\begin{tabular} {|c|c|c|} \hline\hline
Dimensioned units & Dimensionless units                         & Dimensionless value  \\  \hline
t(1 secs)        & $\overline{t}=\frac{t}{m^{0.5}U_0^{-0.5}k^{-1}}$  &  42472.13 \\  \hline
C(1 pFarad)      & $\overline{C}=\frac{C}{m^{0.5}kU_0^{0.5}}$        &   0.4264 \\  \hline
R(1 $\Omega $)   & $\overline{R}=\frac{R}{U_0^{-1}k^{-2}}$            &  0.000099602  \\  \hline
Voltage(1 Volt)  & $\overline{V}=\frac{V}{k^{-1}}$            &   1 \\  \hline
Current(1 mA)    & $\overline{I_0}=\frac{I_0}{U_0k}$            &  10.04  \\  \hline\hline
\end{tabular}
\label{table1}
\end{table*}
\end{center}

The electronic circuit (Fig. 1) is a weakly damped periodically forced nonlinear (feedback) oscillator that simulates a second order ordinary differential equation giving the solution in the form of an output voltage, $V_{out}(t)$, in response to an input current, $I_{inp}(t)$. Note that our drive frequencies are not very different from the characteristic small-amplitude natural frequency of the oscillator. For a periodic $I_{inp}(t)$ of frequency $f$, one is expected to obtain a periodic $V_{out}(t)$ of the same frequency given the parameter $\gamma$ suitably fixed. In the experiment, we choose a suitable frequency $f=f_f$ for which the amplitude of $V_{out}(t)$ is maximum. We set this frequency $f=f_f$ and the amplitude of the input periodic current $I_0=0.2$ (or $V_{in}^0\approx$ 202mVpp corresponding to the value of different parameters given in Fig. (1)) in the waveform generator and we let $V_{out}(t)$ to oscillate. We then switch off the input periodic current $I_{inp}(t)$. Since the oscillator is underdamped, $V_{out}(t)$ oscillates freely with an exponentially diminishing amplitude \cite{Johannessen}, as a solution of the equation:

\begin{equation}
\frac{d^{2}V_{out}}{dt^{2}}=-\left\{\frac{R_2C_1}{R_B}+C_3+\frac{R_2C_2}{R_A}\right\}\frac{dV_{out}}{dt}-
\sin(V_{out})
\label{eqn5}
\end{equation}

Therefore, for our final free oscillation experiment we use the same electronic circuit shown by the block diagram in Fig. \ref{fig1} with the drive $I_{inp}(t)$ switched off giving the solution of equation (\ref{eqn5}). The circuit essentially consists of two integrator segments, the output of one is fed as input to the second at point B through a resistor $R_2$ and two negative feedbacks as input at point A. One (upper) feedback simulates the damping term $\gamma\frac{dV_{out}}{dt}$ through the capacitor $C_3$. The other feedback gives the sinusoidal force term $\sin(V_{out})$ derived from the periodic potential. Of course, the equation (\ref{eqn3}) satisfied at point A assumes the components to be ideal. However, the characteristic parameter values of the real components may differ slightly from the ideal values and hence, for example, the value of $\gamma$ may need adjustment. Also, the other feedback term representing sinusoidal force $\sin(V_{out})$ to be evaluated instantaneously of the continuously changing argument $V_{out}$ needs careful consideration.

The IC AD534 gives the sine of the input signal and from the datasheet, it works according to the equation:
\begin{equation}
V_{sine}=10\sin(\frac{\pi}{2} \times \frac{V_z}{10})
\label{sineconverter}
\end{equation}
where $V_{sine}$ is the output signal from the sine converter and $V_z$ is the input signal to the sine converter, where $V_z$ can go from $-10$V to $+10$V. However, when we use the circuit design as specified in the datasheet, we find that the IC works best only when the input signal is in the range from 0V to 1.165V out of the maximum 10V as specified in the datasheet. Therefore, we modify the parameters related to AD534 so that we can go beyond 1.165V. We use trial and error method and arrive at a conclusion that, for different combinations of the parameters related to AD534, the output from the sine converter should be of the form
\begin{equation}
V_{sine}=4.6\sin(\frac{\pi}{2} \times \frac{V_z}{5})
\label{modifiedsineconverter}
\end{equation}
where $V_z$ can go from -5V to +5V. We found that even if $V_z$ goes to $\pm$5V, the AD534 still gives approximately the required output signal $V_{sine}$. Therefore, we have a sine converter where the argument $\theta = ({\pi}/{2}) \times ({V_z/}{5})$  is in the range [$\frac{-\pi}{2}, \frac{\pi}{2}$].

\begin{figure}[htp]
\centering
\includegraphics[width=9cm,height=6cm]{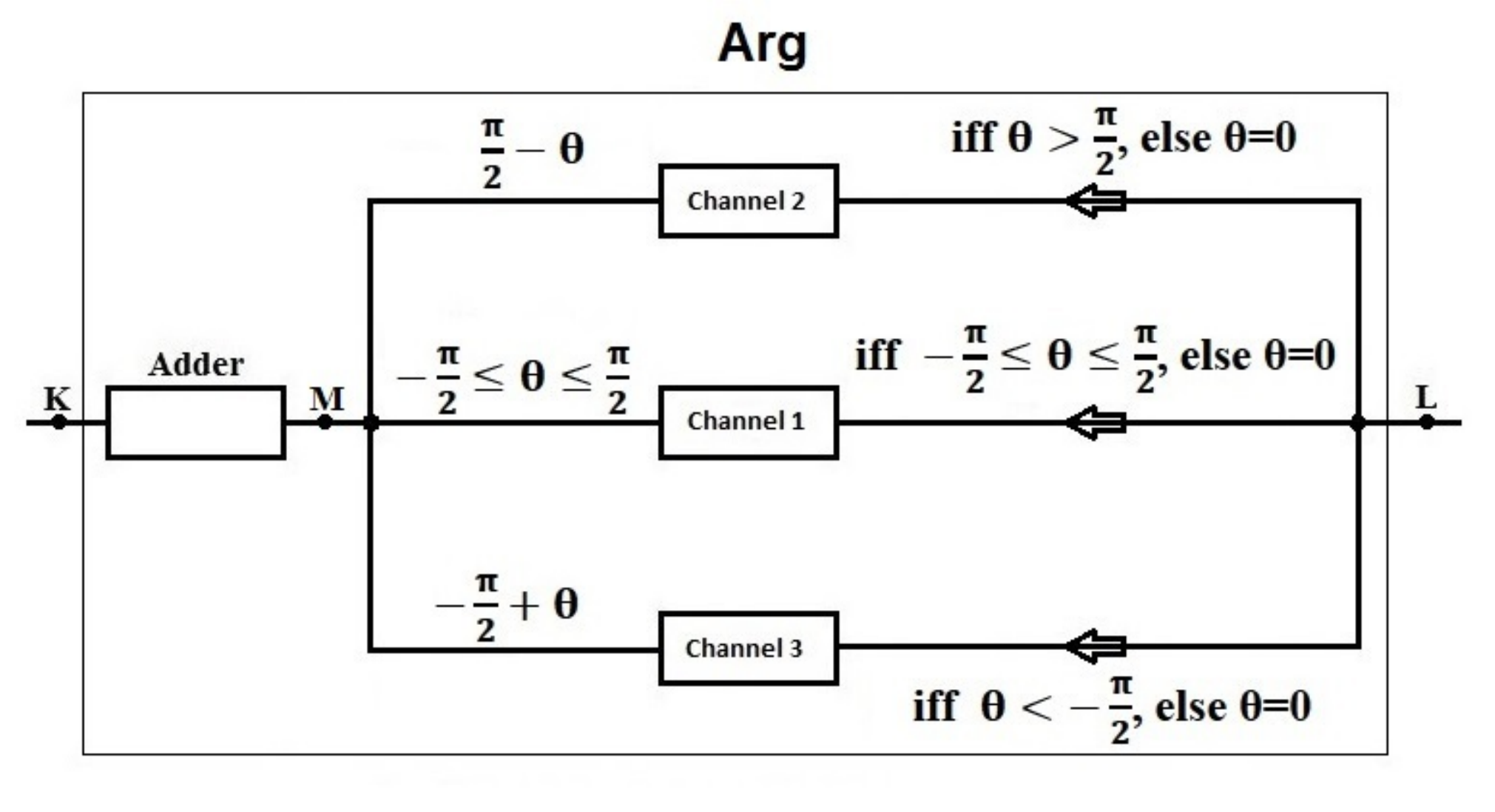}
\caption{Block diagram for the argument of sine to get the full potential}
\label{fig2}
\end{figure}

The sine converter block AD534 of Fig. 1 is tasked to obtain $\sin(\theta)$, $\theta=V_{out}$ which may take values larger than $\pm\frac{\pi}{2}$. However, as mentioned above, it was found that using only the block AD534, without the block [Arg], the sine converter converts only roughly upto a range of argument, $-\pi/2 <\theta <\pi/2$ in dimensionless units. Therefore, we have augmented the feedback segment by adding an extra circuit block [Arg], as shown in the block diagram of Fig. \ref{fig2}, in order to get the full potential, $-\cos(\theta)$, with  $-\pi\leq\theta < \pi$. Noting that the range of argument $-\pi/2\leq\theta\leq\pi/2$ covers the entire range of values of $-1\leq\sin(\theta)\leq1$, we need only to shift the arguments $\theta<-\pi/2$ and $\theta > \pi/2$ appropriately into the range $-\pi/2\leq\theta\leq\pi/2$ using the usual trigonometric rules. In the following we provide a detailed explanation of the circuit block [Arg]. 

\begin{figure}[htp]
\centering
\includegraphics[width=17cm,height=11cm]{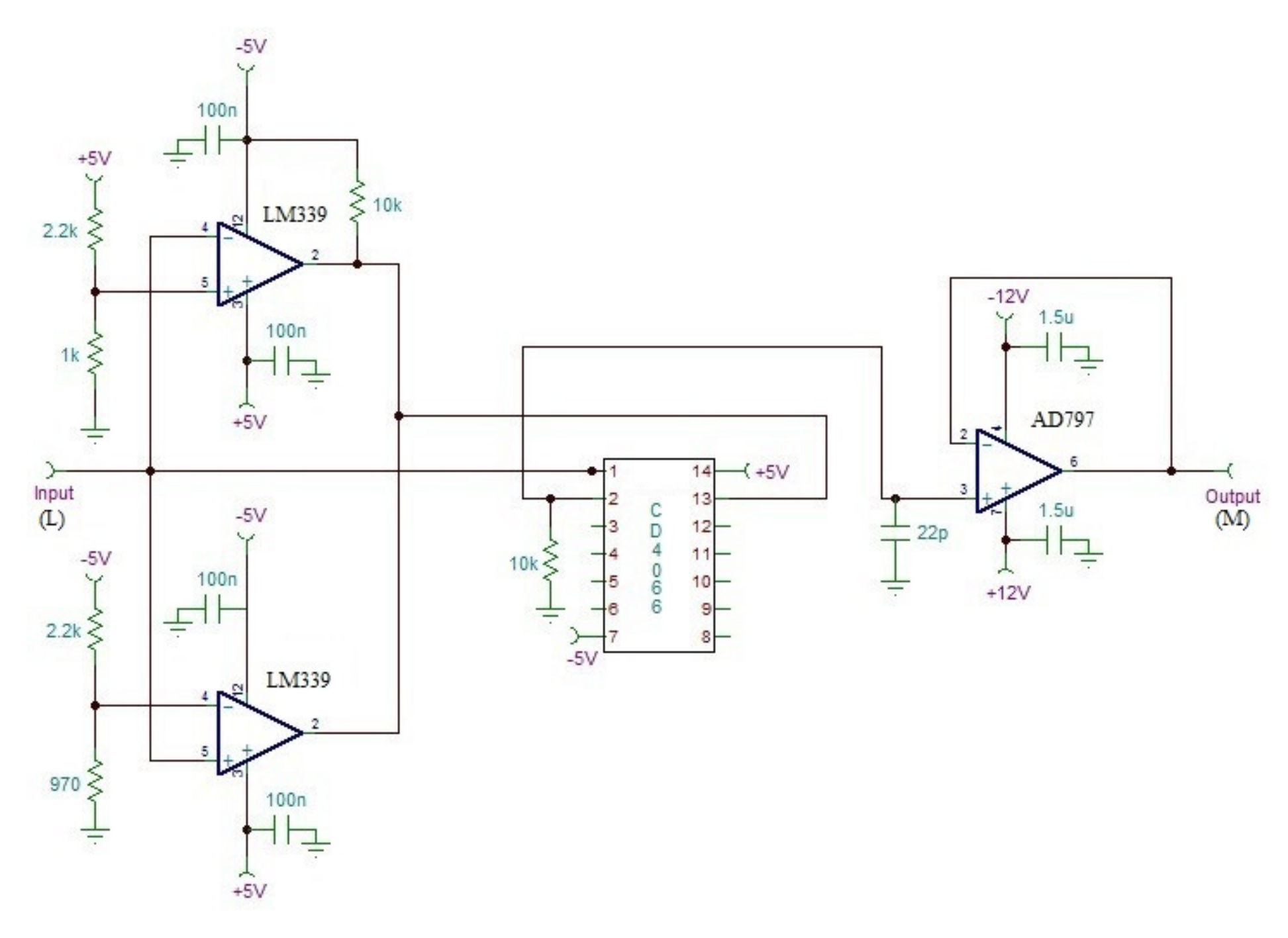}
\caption{Schematic diagram representing Channel 1 from Fig. 2.}
\label{fig2a}
\end{figure}

\begin{figure}[htp]
\centering
\includegraphics[width=23cm,height=11cm,angle=-90]{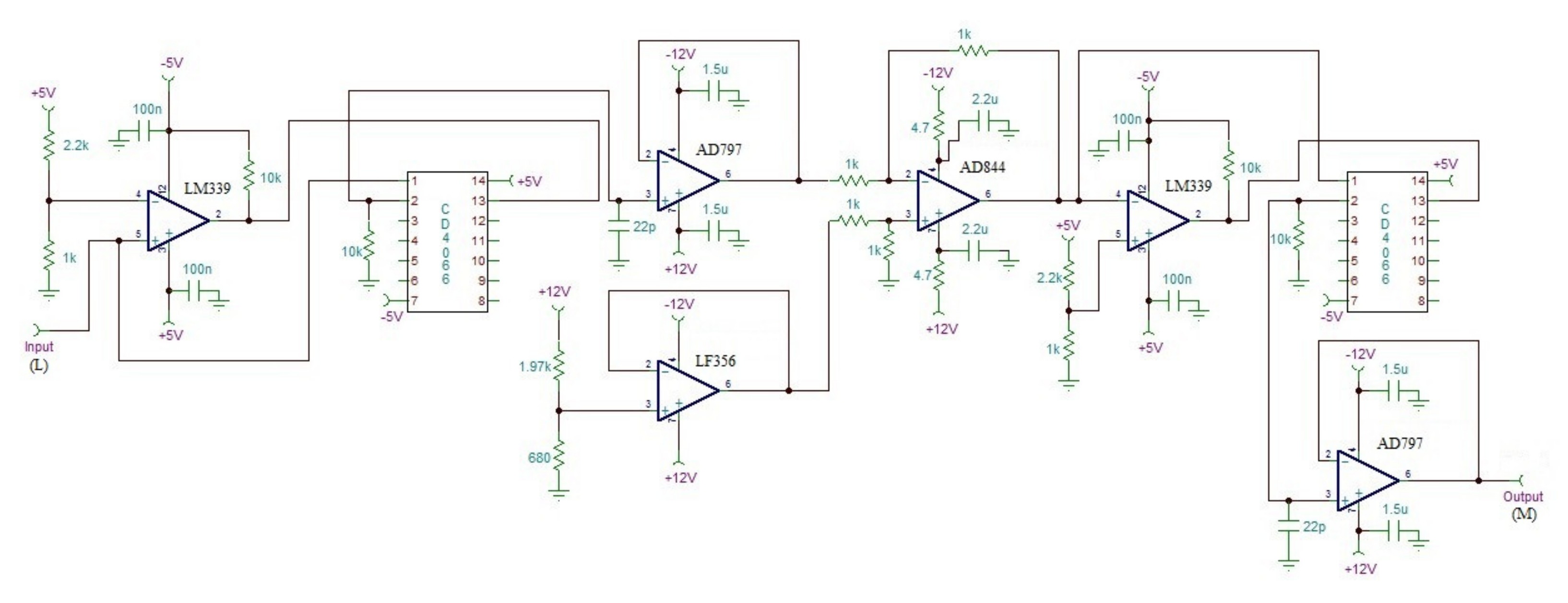}
\caption{Schematic diagram representing Channel 2 from Fig. 2.}
\label{fig2b}
\end{figure}

\begin{figure}[htp]
\centering
\includegraphics[width=23cm,height=11cm,angle=-90]{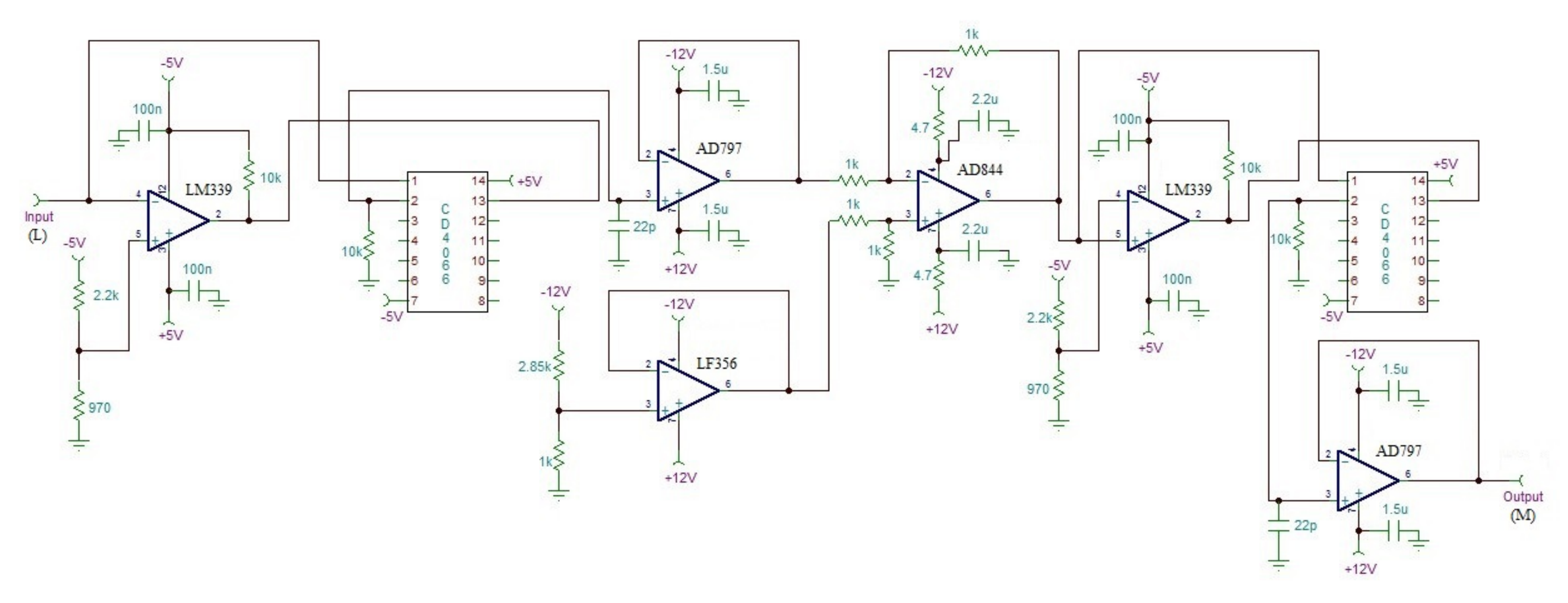}
\caption{Schematic diagram representing Channel 3 from Fig. 2.}
\label{fig2c}
\end{figure}

\begin{figure}[htp]
\centering
\includegraphics[width=9cm,height=6cm]{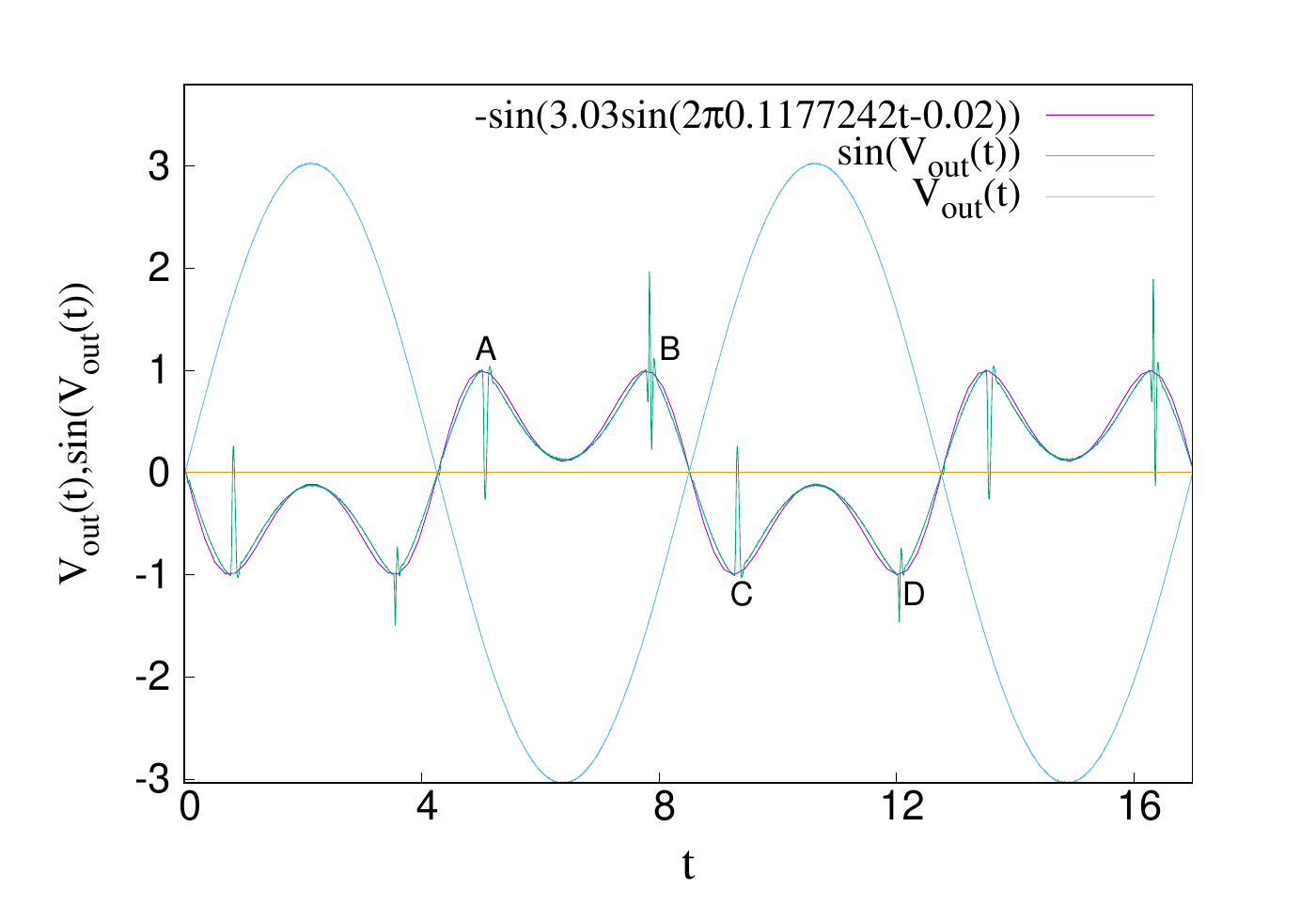}
\caption{A plot between the input signal to the sine converter ($V_{out}(t)$) and the output signal from the sine converter $(\sin(V_{out}(t)))$ along with the fitted curve $-\sin(3.03\sin(2\pi 0.1177242 t-0.02))$.}
\label{fig2d}
\end{figure}

In the circuit shown in Fig. \ref{fig2}, the schematic diagram for the channels 1, 2 and 3 are given in Figs. \ref{fig2a},  \ref{fig2b} and \ref{fig2c}, respectively. The point L, shown in Figs. 2-5, serves as the common input point to all the channels 1, 2, and 3. Similarly, the point M serves as the common output point of all the channels. The input (i.e., the argument $\theta=V_{out}(t)$) can have any value in the range: $-\pi\leq\theta\leq \pi$. The schematic shown in Fig. \ref{fig2a} allows the input signal to go through unchanged only when the input signal is in the range $-\pi/2< \theta < \pi/2$ and beyond this range, the output signal is set equal to zero. The argument $\theta'$, which is the output signal at point M in Fig. \ref{fig2a}, therefore remains unchanged, $\theta'=\theta$. The schematic shown in Fig. \ref{fig2b} accepts the input signal to be converted only when the input signal is in the range $\pi >\theta > \pi/2$ and when $\theta$ is not in this range the output signal is set zero. This Channel 2 converts the allowed value of $\theta$ to $\theta'=(\pi-\theta)$ so that the output signal ($=\theta'$) from this schematic (at point M) is either in the range $0<\theta'<\pi/2$ or $\theta'=0$. Similarly, the schematic shown in Fig. \ref{fig2c} accepts the input signal to be converted only when the input signal is in the range $-\pi < \theta <-\pi/2$ and beyond this range, the output signal is set zero. This Channel 3 converts the allowed value of $\theta$ to $\theta'=-\pi-\theta$ so that the output signal from this schematic (at point M) is either in the range $0>\theta'>-\pi/2$ or $\theta'=0$. The output from these three channels are then added using an adder shown in Fig. \ref{fig2}. The output from the adder becomes the input argument to the sine converter at point K. Thus, using these three complementary channels simultaneously, we have an input signal (argument) suitable for the sine converter in the whole range $-\pi\leq\theta < \pi$. Therefore, in Fig. 1, the transformed voltage $V_K$ at point K appears in the limited range of $-\pi/2\leq V_{K} < \pi/2$ just appropriate as input to the IC AD534 for the whole range of $-\pi\leq V_{out}(t) < \pi$.

We have used an inverter with gain=$\frac{10}{\pi}$ after point K so that from Eqn. \ref{modifiedsineconverter}, $V_z$ is given by $V_z=V_{K}\times\frac{10}{\pi}$. This is done in order for the voltage $V_K$ to be the right argument in the sine converter. In the circuit of Fig. 1, we have also used an inverter with gain=$\frac{1}{4.6}$ in order to compensate for the scale factor=4.6 given in Eqn. \ref{modifiedsineconverter}. Hence, the voltage at point P is $-\sin(V_{out})$, as envisaged.

In order to verify the efficacy of the module [Arg] described above we plot, in Fig. \ref{fig2d}, the input voltage to the module [Arg] (represented by $V_{out}(t)$ in Fig. 1) and the final signal obtained from the sine converter at point P.  The output signal is fitted with a function $-\sin(3.03\sin (2\pi 0.1177242t-0.02))$ where the input signal has an amplitude 3.03V and a frequency=5000Hz. In terms of dimensionless units, for the parameters given in the caption of Fig. \ref{fig1}, the amplitude of the input signal is 3.03 and the frequency is 0.1177242. We see that the sine of the input signal matches quite well with the theoretical curve except for the spikes at positions close to the argument $\theta=\pm \pi/2$ where the channels switch. The spikes in the output signal from the sine converter, $\sin(V_{out}(t))$, are because of the switchings from one channel to the other. For example, when the signal $V_{out}(t)$ goes from $<\pi/2$ to $>\pi/2$, the channel switches from channel 1 to channel 2 and since the comparator LM339 in Fig. 3-5, has a response time of $\approx 1\mu sec$, it takes $\approx 2\mu sec$ for the channel 2 to become fully functional. At $\approx \pi/2$, channel 1 tries to go to zero where the response time is $\approx 1\mu sec$ but channel 2 wants to change the input signal to $\pi-\theta$ where its response time is again $\approx 1\mu sec$ and hence we see an abrupt dip and rise (with a total response time of $\approx 2\mu sec$) in the output signal before channel 2 activates as shown by the point A in Fig. \ref{fig2d}. Similarly, for the spikes at points B, C and D. The frequencies used in our experiment are around 5300Hz (time period $ \approx 188.7\mu sec$) and since the total response time of a spike is $\approx 2\mu sec$, these spikes do not affect our overall experimental results. 

As a small remark, when the amplitude of the output signal $V_{out}(t)$ is very close to $\pi/2$, that is, when $\sin(V_{out}(t))$ has a flattened profile the system cannot unambiguously decide whether to go to the channel 1, 2, or 3 of the circuit shown in Fig. \ref{fig2}. Because of this, the output at point M, for this particular amplitude, fluctuates. However, this is not a common occurrence when the amplitude of $V_{out}(t)$ is slightly $>+\pi/2$ (or $<+\pi/2$) or slightly $<-\pi/2$ (or $>-\pi/2$) by even a small finite value, this problem of fluctuation disappears. Therefore, this special rare occurrence does not concern us in our experiment.

\maketitle
\section{Experimental Results}

As mentioned earlier, for a given damping coefficient $\gamma$, initially the system is periodically driven with a frequency so that we can have a maximum amplitude of response $V_{out}(t)$. For all values of $\gamma$ we take the drive current amplitude $I_0=(V_{in}^0R_4/V_0R_1)$ same and $\approx 0.2$ and choose the frequency. For example, for $\gamma=0.0795$ the appropriate frequency was found to be 5.3 kHz or the angular frequency $\omega=0.784$ in dimensionless units. The external drive is then switched off at an instant when the response $V_{out}$ was close to maximum and the free oscillation $V_{out}(t)$ was measured. Note that all the data points, for example, $V_{out}$, $t$, etc. are obtained from the oscilloscope readings which are subsequently converted into dimensionless units and presented as measured values in our results. Figs. 7, 9-11, show our experimental results for various measured values of $\gamma$, all in dimensionless units.

\subsection{Variation of amplitude with time}

\begin{figure}[htp]
\centering
\includegraphics[width=17cm,height=8cm]{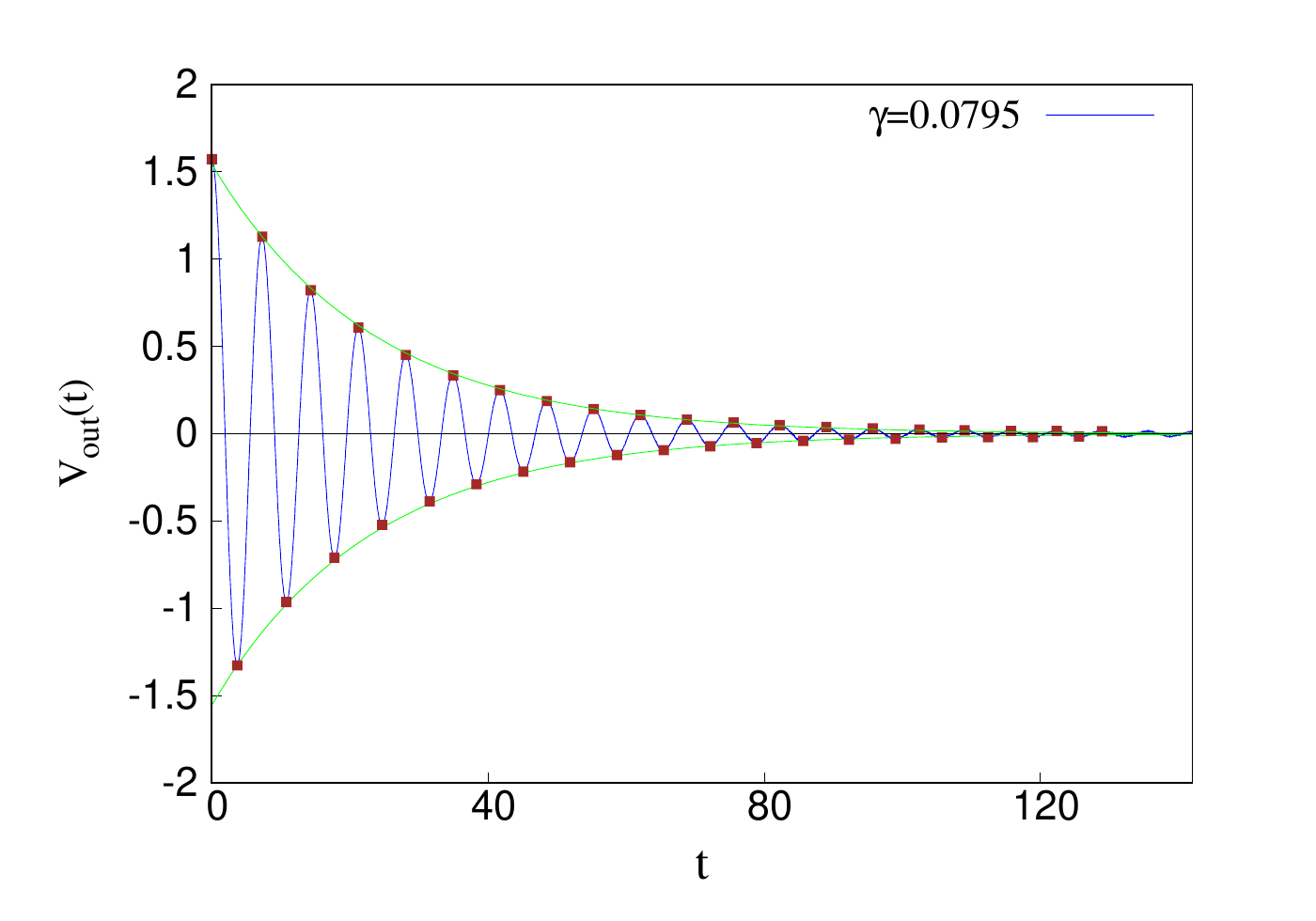}
\caption{Plot of the free oscillation of the output signal voltage ($V_{out}(t)$) (blue line) as a function of time when $\gamma =0.0795$.}
\label{fig3}
\end{figure}

Figure \ref{fig3} shows the free oscillation of the output signal voltage ($V_{out}(t)$)(blue line) as a function of time for $\gamma =0.0795$. While saving the data from the oscilloscope, the free oscillation of $V_{out}(t')$ can start at any instant of time $t'$ not necessarily equal to 0. However, in Fig. \ref{fig3}, we have shifted the origin of time so that the free oscillation of $V_{out}(t)$ starts at the shifted time $t=0$, either from a maximum or a minimum of $V_{out}$ as desired. The measured maxima and minima of the oscillation are shown by the brown points. The maxima of the plot are fitted with the equation $1.57448e^{-\frac{\gamma'}{2}t}$ and the minima are fitted with $-1.32317e^{-\frac{\gamma' }{2}t}$ where $t=t'-t_+$ or $t=t'-t_-$, according as whether we consider, respectively, the maxima or the minima of $V_{out}$, for the effective damping coefficient $\gamma'\approx 1.08\gamma$.   Here, $t'=t_+=4092.12211$ represents the instant at which the observation of free oscillation of $V_{out}$ starts at a maximum (=1.57448) and $t'=t_-=4095.80019$ represents the instant of the subsequent minimum (= -1.32317) of $V_{out}$. For both the fitting functions, the plots are shown by the green lines with the  general equation $V_{out}^me^{-\frac{\gamma' }{2}t}$, where $V_{out}^m$ is the extremum value of $V_{out}$ at $t=0$. We see that the fits are quite good. Note that $V_{out}^m$, $t_+$ and $t_-$ can be different for different sets of experiment. The exponentially decreasing amplitude of the underdamped system reaches the minimum measurable value after a certain time (say $t_d$). As we increase the $\gamma $ value, we see that the number of measurable cycles of free oscillation also decreases and the corresponding time duration $t_d$ upto which the free oscillations can be measured unambiguously shortens, as expected.

\begin{figure}[htp]
\centering
\includegraphics[width=13cm,height=7cm]{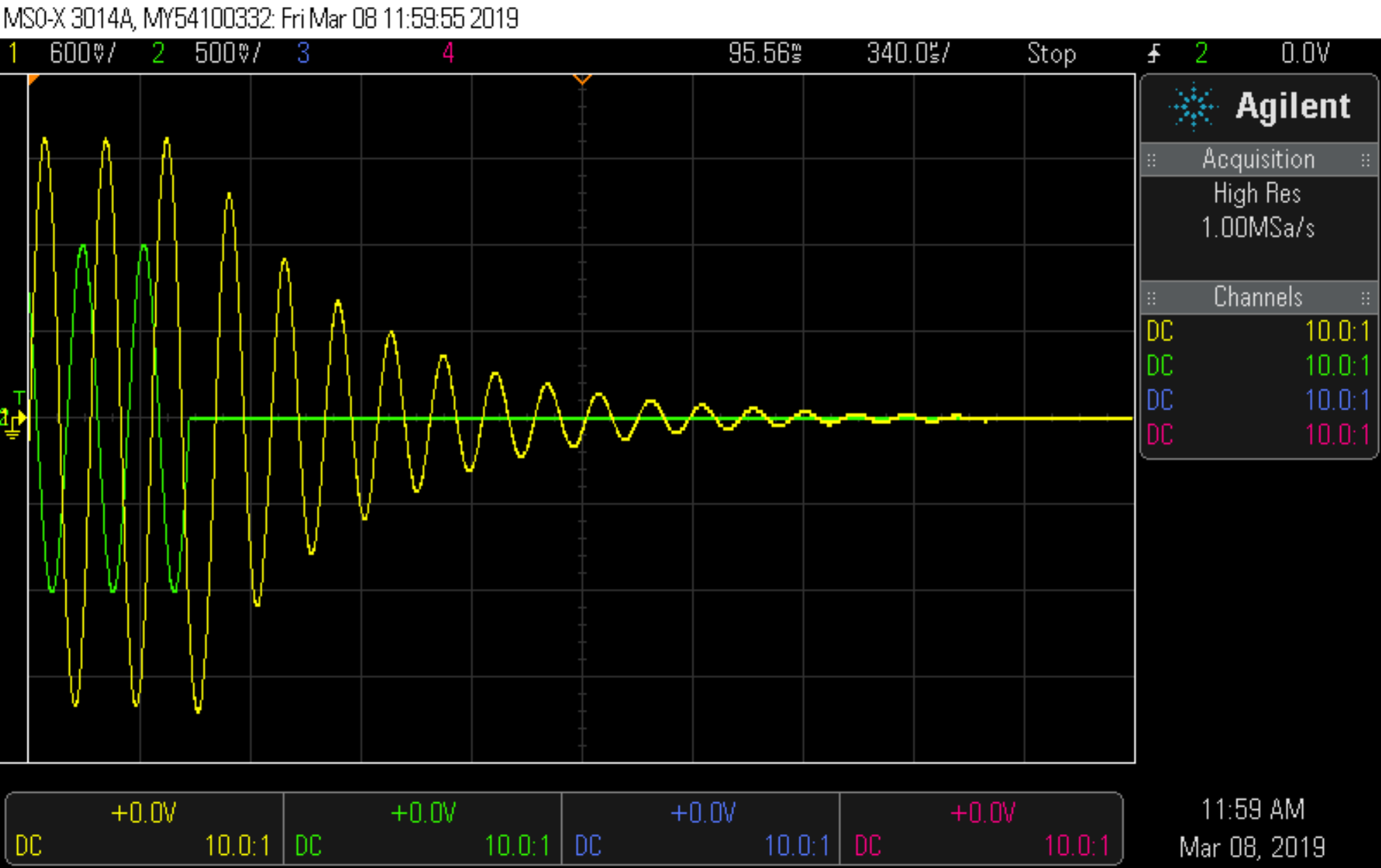}
\caption{Plot of the free oscillation of the output signal voltage ($V_{out}(t)$) (yellow line) and the input signal voltage ($V_{in}(t)$) (green line) as a function of time when $\gamma =0.0795$.}
\label{fig3i}
\end{figure}

Fig. \ref{fig3i} shows the free oscillation of the output signal voltage ($V_{out}(t)$) along with the input signal voltage ($V_{in}(t)$) as a snapshot from the oscilloscope, illustrating the free oscillations shown in Fig. \ref{fig3}. The physical quantities in Fig. \ref{fig3i} are in units of volt and sec whereas in Fig. \ref{fig3}, we have used dimensionless units.

\begin{figure}[htp]
\centering
\includegraphics[width=17cm,height=8cm]{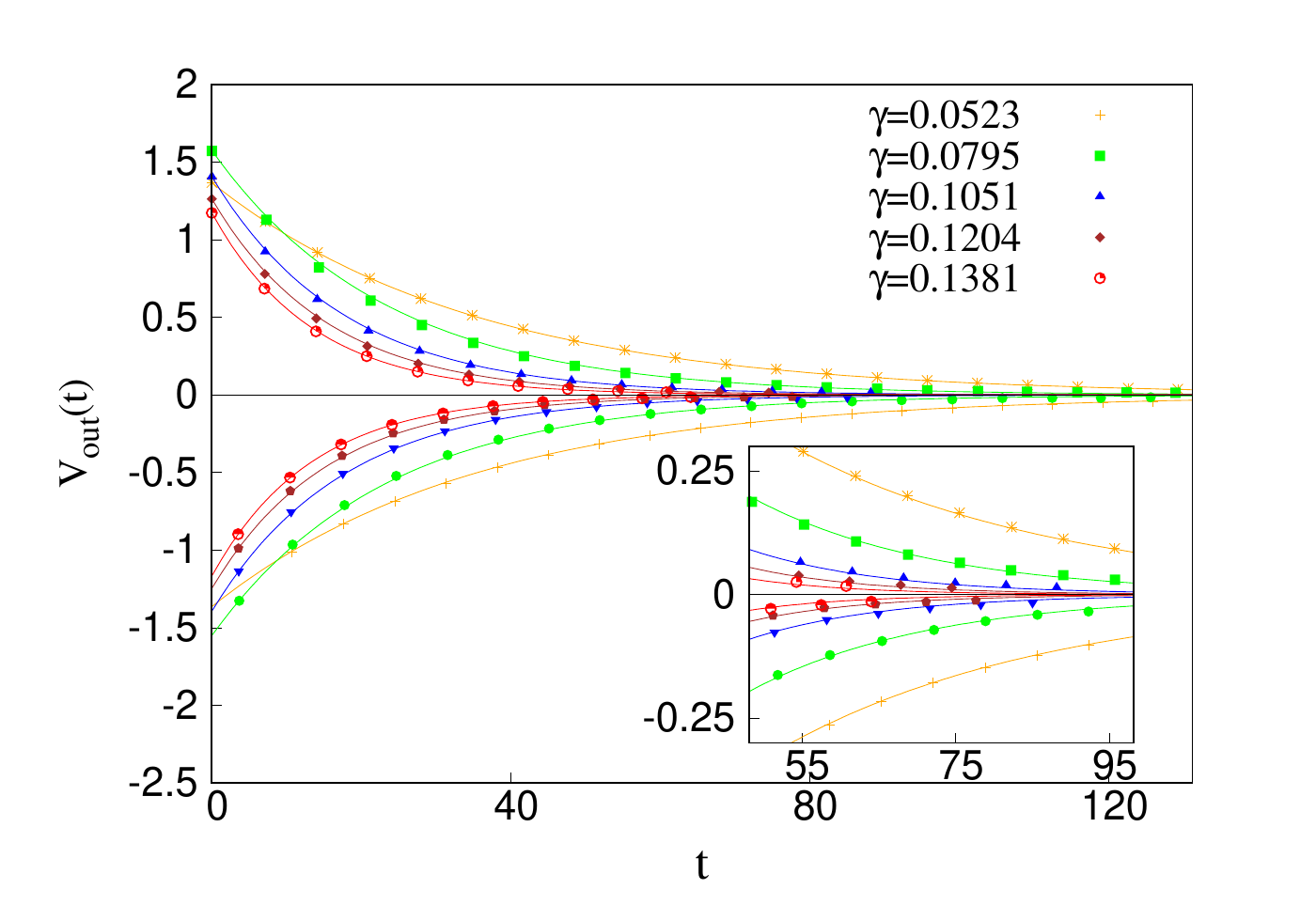}
\caption{Plot of the measured peak of the amplitudes of the output signal voltage ($V_{out}(t)$) as a function of time for $\gamma =0.0523,~0.0795,~0.1051,~0.1204$ and $0.1381$. The measured peak of the amplitudes is shown by the points and the fitted curve ($\approx V_{out}^me^{-\frac{\gamma }{2}t}$) is shown by the thick lines.}
\label{fig4}
\end{figure}

In Fig. \ref{fig4}, we show how the amplitudes of oscillation decay with time $t$ for several values of $\gamma=0.0523,~0.0795,~0.1051,~0.1204$ and $0.1381$, just as in Fig. \ref{fig3} for the lone $\gamma =0.0795$. As noted earlier, the measured times $t_+,~t_-$ and $t'$ from the oscilloscope have different values for each $\gamma$ and for each set of experiment. However, when the figures are plotted as a function of $t=t'-t_+$ or $t=t'-t_-$ (instead of as a function of $t'$), all the curves begin with $t=0$ (instead of beginning with $t'=t_+$ or $t'=t_-$), for all $\gamma$. The curves are thus automatically time shifted appropriately. Of course, the curves begin with different amplitudes $V_{out}^m$ as shown in the figure. All the fitted curves have the same form $V_{out}^me^{-\frac{\gamma' }{2}t}$. The inset of Fig. \ref{fig4} shows the magnified picture of the curves at large $t$ values when the amplitude is small. The curve fittings appear to be good even for these large times. However, a semilogarithmic plot should give a better check.

\begin{figure}[htp]
\centering
\includegraphics[width=17cm,height=8cm]{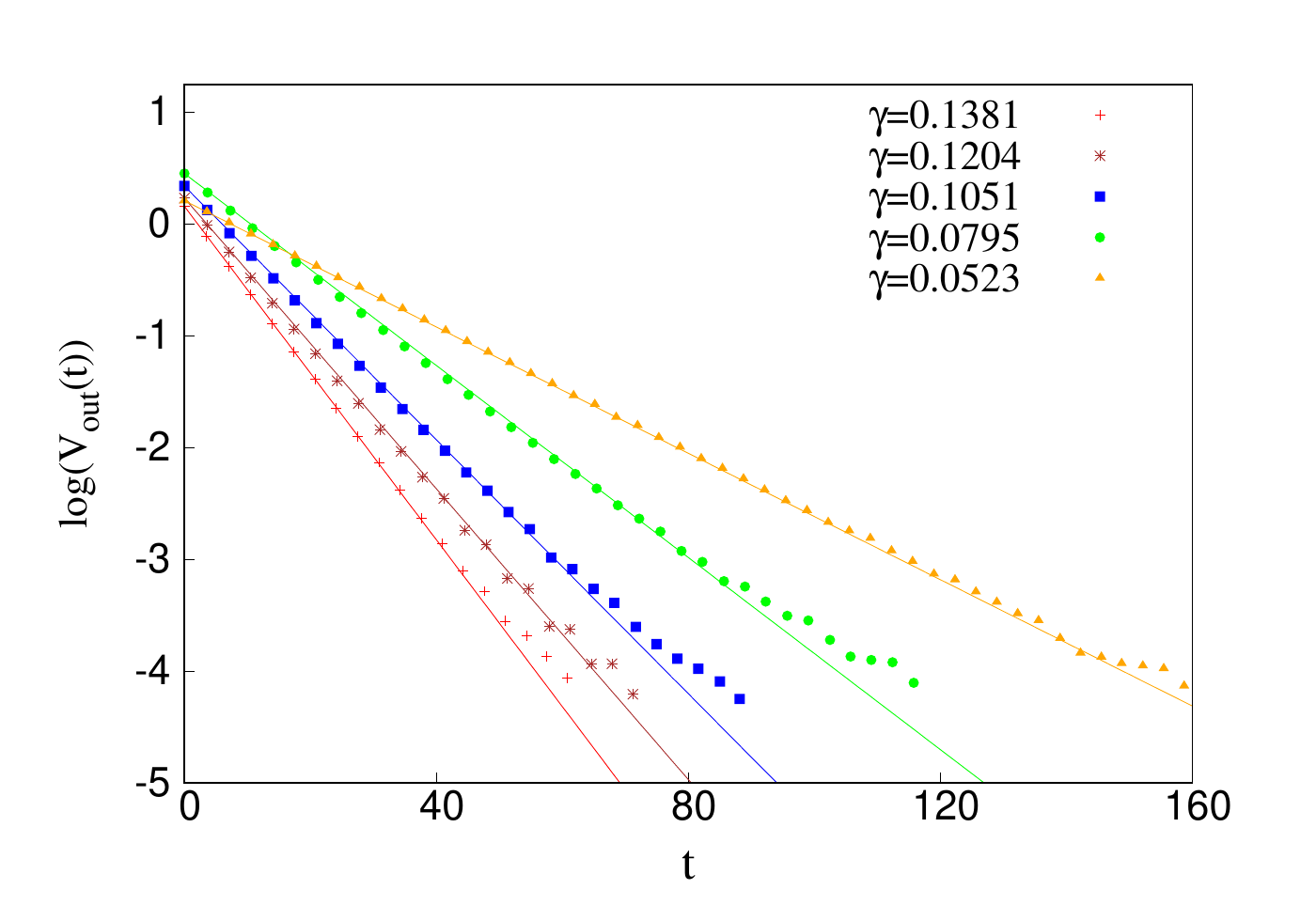}
\caption{Semilogarithmic plot of the amplitude of Fig. (\ref{fig4}) for $\gamma =0.0523,~0.0795,~0.1051,~0.1204$ and $0.1381$ as a function of time.}
\label{fig5}
\end{figure}

The time varying amplitude shown in Fig \ref{fig4} for different $\gamma $ values are replotted in the semilogarithmic graph of the magnitudes of both the maxima and minima of $V_{out}$ as a function of time $t$ in Fig. \ref{fig5}. The curves are then fitted with  straight lines having slopes equal to $-0.54\gamma$ (= $-\frac{\gamma'}{2}$). The fittings are quite good in the entire range of $t$ except for the large $t$ values or for small amplitude values. In the small amplitude range also one can fit roughly by straight lines but with smaller slopes than $\frac{\gamma'}{2}$. Theoretically, to conform to the damped harmonic oscillator case, the slope of the fitted curve even for the small $t$ range should have been $\frac{\gamma}{2}$ in place of $\frac{\gamma'}{2}$. The discrepancy could be because of approximate values of used parameters of the components in the expression for $\gamma =\{\frac{R_2C_1}{R_B}+C_3+\frac {R_2C_2}{R_A}\}(R_4/m)^{0.5}$. As noted earlier, the IC's are not ideal and wires do have a parasitic capacitance and resistance, and hence the calculated values of a damping coefficient $\gamma$ may not be the same as that actually appearing in the circuit. Also, the differing slope of the fitted curves in the two regions of small and large amplitude could be because of variation of frequency of oscillation as a function of its amplitude unlike the amplitude independent constant frequency in the case of damped simple harmonic oscillators.

\begin{figure}[htp]
\centering
\includegraphics[width=17cm,height=8cm]{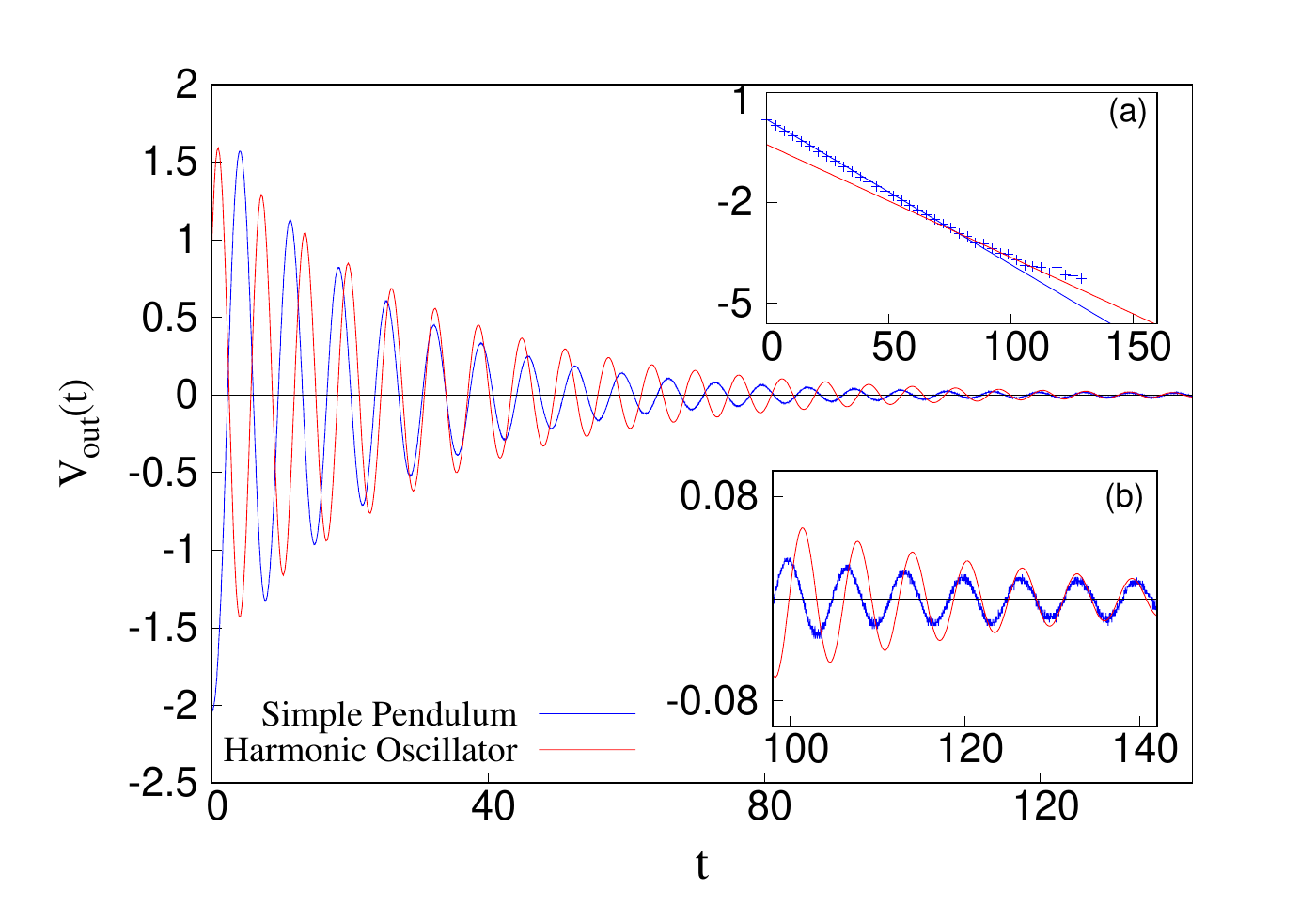}
\caption{Plot of the free oscillation of the output signal voltage ($V_{out}(t)$) (blue line) as a function of time when $\gamma =0.0795$ for a simple pendulum where we have fitted with a damped simple harmonic motion (red line).}
\label{fig3a}
\end{figure}

In Fig. \ref{fig3a} we compare a sample trajectory $V_{out}(t)$ for $\gamma =0.0795$ (same as the plot shown in Fig. \ref{fig3}) and a theoretically calculated trajectory of a damped harmonic oscillator. The frequency of the harmonic oscillator is chosen to fit it closely with $V_{out}(t)$ at small amplitudes as in the inset Fig. \ref{fig3a}b and the damping coefficient $\gamma=0.42\times 0.0795$ (red line) obtained from the fit shown in the inset Fig. \ref{fig3a}a. From the inset Fig. \ref{fig3a}a we see that the effective $\gamma$ value changes from $0.54\times 0.0795$ (blue line) to $0.42\times 0.0795$ (red line). From the two trajectories, one can clearly see that the frequency of oscillation in the case of a sinusoidal potential differs from that in the case of a simple harmonic motion as the amplitude increases. 

\subsection{Change in frequency of free oscillation with amplitude}

\begin{figure}[htp]
\centering
\includegraphics[width=17cm,height=8cm]{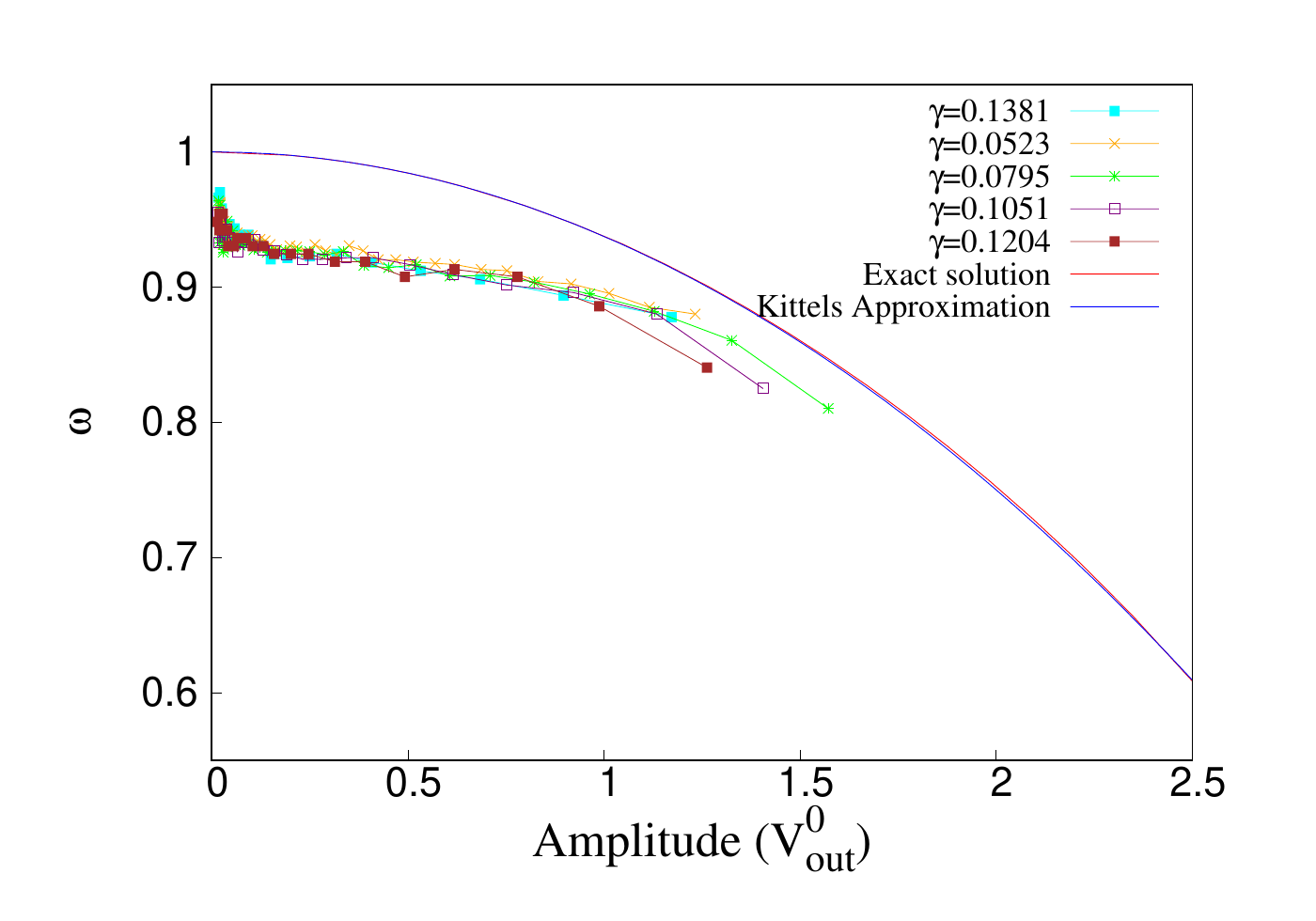}
\caption{Plot of the angular frequency  of free oscillation ($\omega $) as a function of the measured peak of the amplitudes of the output signal voltage ($V_{out}(t)$) for $\gamma =0.0523,~0.0795,~0.1051,~0.1204$ and $0.1381$.}
\label{fig6}
\end{figure}
Figure \ref{fig6} shows the measured angular frequency of free oscillation of the analog circuit model simple pendulum as a function of the amplitude for $\gamma =0.0523, 0.0795, 0.1051, 0.1204$ and $0.1381$. From the plot, we see that the free oscillation starts with a particular initial frequency of oscillation which increases as the amplitude of the free oscillation decreases. This is to be compared with the amplitude independent frequency ($\omega _1=\sqrt{1-\frac{\gamma ^2}{4}}$) of oscillation of a damped harmonic oscillator. In Fig. \ref{fig6}, the angular frequency of oscillation for a particular $\gamma$ value does not differ too much from other $\gamma$ values. Here we have compared the experimental results with the oscillation of a simple pendulum using the approximate expression used by Kittel et al \cite{Kittel}

\begin{equation}
\frac{\omega_1 }{\omega_0 }\approx 1-\frac{x_0^2}{16}
\end{equation}

and also with the exact solution \cite{Sommerfeld,Belendez2011}

\begin{equation}
\frac{\omega_1 }{\omega_0 }\approx \frac{\pi}{2K(k)}
\end{equation}

where

\begin{equation}
K(k)=\int_{0}^\frac{\pi}{2} \frac{d\phi }{\sqrt{1-k\sin^2 \phi }}
\label{elliptic}
\end{equation}

and

\begin{equation}
k=\sin^2\frac{x_0}{2}
\end{equation}
plotted in Fig \ref{fig6}, using the standard tables \cite{Abramowitz} for the elliptic integrals (\ref{elliptic}). Here, both the plot for the exact solution and the approximate solution is for large amplitude oscillation and in the absence of damping.  When these two results are compared with the experimental results, we see that for a particular amplitude, the measured angular frequency of oscillation is smaller by about 8\% compared to the theoretically obtained frequency \cite{Sommerfeld,Belendez2011}. As mentioned by Squire \cite{Patrick}, the angular frequency of oscillation of a simple pendulum in the absence of damping should be very close to that in the presence of damping. This simply shows that our measured frequencies have an error of about 8\%. However, the trend of variation of the measured frequency is qualitatively similar to the theoretical one.

\section{Discussion and Conclusion}

We have presented solutions to the differential equation of motion describing the force-free oscillations of a damped simple pendulum using an analog electronic circuit. Naturally, the pendulum is highly nonlinear and oscillates under the full sinusoidal potential force field. We have presented the details of how to obtain the sinusoidal force field.

Since exact analytical solutions to the motion of damped simple pendulum are not available, it is worth examining experimentally the expected approximate solutions. Our experimental results show correct qualitative trends when compared with the exact analytical results for the undamped simple pendulum. The results also illustrate how oscillations of a simple pendulum differ from those of a harmonic oscillator. The experiment could be useful and  educative at the undergaduate level to make a clear distinction between a pendulum with a large amplitude of oscillation and the one usually learnt with a small amplitude approximation.

In our model, the damping is taken to be proportional to the instantaneous velocity. The coefficient of damping is considered constant and, in our circuit model, determined by the parameters of various components used in the circuit. When this calculated damping coefficient is compared with the damping coefficient calculated from the usual amplitude decay exponent we find a discrepancy of about 8\%. There could be many factors causing this discrepancy and the nonideal nature of components used in the circuit could be one. Also, the measured amplitude dependent frequency of oscillation differs from the theoretically calculated one by a similar percentage. At present, we do not have any plausible explanation for this discrepancy, however.

As explained earlier, Squire pointed out that in case of a real pendulums \cite{Patrick}, at least two damping terms are generally needed. For a rigid pendulum, apart from the damping term which is linear in velocity, air drag is always present contributing to a damping term quadratic in velocity. However, in our model we consider only the linear damping term to keep the circuit as simple as possible so that the experiment can be replicated easily even in undergraduate teaching laboratories.

\end{document}